\documentclass[prl,twocolumn,superscriptaddress,amsmath,amssymb]{revtex4}
\usepackage{graphicx}
\usepackage{bm}
\usepackage{amsmath,amssymb}
\usepackage{color}
\usepackage{ascmac}

\begin{document}
\title{Secure entanglement distillation for double-server blind 
quantum computation}  
\author{Tomoyuki Morimae}
\email{morimae@gunma-u.ac.jp}
\affiliation{ASRLD Unit, Gunma University,
1-5-1 Tenjin-cho Kiryu-shi Gunma-ken, 376-0052, Japan}
\affiliation{Department of Physics, Imperial College London,
London SW7 2AZ, United Kingdom}

\author{Keisuke Fujii}
\email{keisukejayorz@gmail.com}
\affiliation{The Hakubi Center for Advanced Research, Kyoto University,
Yoshida-Ushinomiya-cho, Sakyo-ku, Kyoto 606-8302, Japan}
\affiliation{Graduate School of Informatics, Kyoto University,
Yoshida Honmachi, Sakyo-ku, Kyoto 606-8501, Japan}
\affiliation{Graduate School of Engineering Science, Osaka University,
Toyonaka, Osaka 560-8531, Japan}
\date{\today}
            
\begin{abstract}
Blind quantum computation is a new secure quantum computing protocol
where a client, who does not have enough quantum technologies at her disposal, 
can delegate her quantum computation to a server, who has a fully-fledged
quantum computer, in such a way that the server cannot learn anything
about client's input, output, and program.
If the client interacts with only a single server,
the client has to have some minimum quantum power, such as
the ability of emitting randomly-rotated single-qubit states
or the ability of measuring states.
If the client interacts with two servers who share Bell pairs
but cannot communicate with each other, the client 
can be completely classical.
For such a double-server scheme, two servers have to share clean
Bell pairs, and therefore the entanglement distillation is necessary
in a realistic noisy environment.
In this paper, we show that it is possible to perform entanglement distillation
in the double-server scheme without degrading the security
of the blind quantum computing.
\end{abstract}

\pacs{03.67.-a}
\maketitle  
A first generation quantum computer will be implemented
in a ``cloud" style, since only limited number of groups,
such as huge industries and governments, will be able to possess it.
When a client uses such a quantum server via a remote access,
it is crucial to protect client's privacy.
Blind quantum computation~\cite{BFK,FK,Barz,Vedran,AKLTblind,topoblind,CVblind,topoveri,MABQC,Sueki,composable,composableMA}
is a new secure quantum computing protocol which
can guarantee the security of client's privacy
in such a cloud quantum computing.
Protocols of blind quantum computation
enable a client (Alice), who does not have enough quantum technologies
at her disposal,
to delegate her quantum computation to a server (Bob), who has
a fully-fledged quantum computer,
in such a way that Alice's input, output, and program are hidden
to Bob~\cite{BFK,FK,Barz,Vedran,AKLTblind,topoblind,CVblind,topoveri,MABQC,Sueki,composable,composableMA}.

The original protocol of blind quantum computation was proposed
by Broadbent, Fitzsimons, and Kashefi (BFK)~\cite{BFK}.
Their protocol uses the measurement-based quantum
computation on the cluster state (graph state) 
by Raussendorf and Briegel~\cite{Raussendorf}.
A proof-of-principle experiment of the BFK protocol
has been also achieved recently with a quantum optical system~\cite{Barz}.
The BFK protocol has been recently generalized to
other blind quantum computing
protocols which use the measurement-based quantum computation
on the Affleck-Kennedy-Lieb-Tasaki (AKLT) state~\cite{AKLT,AKLTblind,Miyake},
the continuous-variable measurement-based quantum 
computation~\cite{CV,CVblind},
and the ancilla-driven model~\cite{Janet,Sueki}.

Since the original BFK protocol was proposed, new protocols have been developed 
in order for blind quantum computation to be more practical.
One direction is making blind protocols more fault-tolerant.
While the BFK protocol, which utilizes the brickwork state,
would be fault-tolerant, its threshold value is extremely small.
The recently proposed topological blind quantum computation~\cite{topoblind}
employs a special three-dimensional cluster state~\cite{Raussendorf_topo}
and allows us to perform topologically protected blind quantum computation
even with a high error probability $0.43\%$ (i.e., fidelity of $99.57\%$) 
in preparations, measurements,
and gate operations.

Another direction is making Alice as classical as possible.
In the above BFK-based protocols, Alice emits randomly-rotated 
single-qubit states, such as single-photon states.
Recently, it was shown~\cite{Vedran} that in stead of single-photon states,
coherent states are also sufficient.
Since coherent states are considered to be more classical than
single-photon states, this result suggests
that Alice can be more classical.

It is also possible to make Alice completely classical:
the double-sever blind protocol was introduced in Ref.~\cite{BFK},
where two Bobs share Bell pairs (but cannot perform classical communication with each other)
and perform computational tasks ordered by Alice's classical message.
The double-server blind protocol is also fault-tolerant,
but Bell pairs of fidelity above 99\% are required even if 
topological blind quantum computation is employed.
Since Bell pairs have to be sent from the third party or Alice herself via public quantum channels,
such an ability to generate high-fidelity Bell pairs or encoding them into 
quantum error correction codes would be too demanding.

In this paper we settle this problem.
We show that it is possible to perform
entanglement distillation in the double-server scheme
without degrading the security of blind quantum computing.
As a result, the required fidelity of the Bell pairs
is improved dramatically to $81\%$,
which is determined by the hashing bound and achieved
by quantum random coding~\cite{Bennett_PRL,Bennett_PRA}.
Since the Bell pair generation of fidelity higher than $81\%$ is
nowadays easily achievable by using, for example, 
parametric down conversion,
the present result is crucial in blind quantum computation 
to make Alice (or the third party) as classical as possible by using practically noisy 
Bell pair sources and quantum channels.

\begin{figure}[htbp]
\begin{center}
\includegraphics[width=0.25\textwidth]{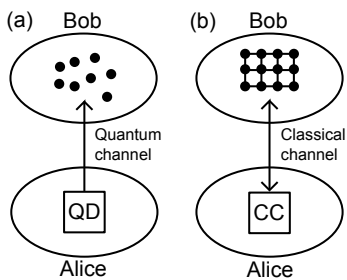}
\end{center}
\caption{
The single-server blind protocol.
(a) Alice sends many single-qubit states to Bob.
QD is a device which emits randomly rotated single qubits.
(b) Bob creates a resource state. Alice and Bob perform
the measurement-based quantum computation through the two-way
classical channel. CC is a classical computer.
} 
\label{single}
\end{figure}

Before proceeding to our main result, 
let us briefly review the BFK blind protocol~\cite{BFK}.
Assume that Alice wants to perform the measurement-based
quantum computation
on the $m$-qubit graph state corresponding to the graph $G$.
The quantum algorithm which Alice wants to run is specified with
the measurement basis $\{|0\rangle\pm e^{i\phi_j}|1\rangle\}$
for $j$th qubit ($j=1,2,...,m$), where
$\phi_j\in\{\frac{k\pi}{4}|k=0,1,...,7\}$.
(Note that such $X-Y$ plain measurements are universal~\cite{Raussendorf}.)
The BFK protocol runs as follows (see also Fig.~\ref{single}):
\begin{itemize}
\item[S1.]
Alice tells Bob the graph $G$~\cite{graph}.
\item[S2.]
Alice sends Bob
$\bigotimes_{j=1}^m|\theta_j\rangle$, where 
$|\theta_j\rangle\equiv|0\rangle+e^{i\theta_j}|1\rangle$
and
$\theta_j$ is randomly chosen by Alice
from $\{\frac{k\pi}{4}|k=0,1,...,7\}$.
\item[S3.]
Bob makes
$|G\{\theta_j\}\rangle\equiv\Big(\bigotimes_{(i,j)\in E}CZ_{i,j}\Big)
\bigotimes_{j=1}^m|\theta_j\rangle$,
where $E$ is the set of edges of $G$
and $CZ_{i,j}$ is the $CZ$ gate between $i$th and $j$th qubits.
\item[S4.]
Alice and Bob now perform the measurement-based quantum computation
on $|G\{\theta_j\}\rangle$ with two-way classical communications as follows:
when Alice wants Bob to measure $j$th qubit ($j=1,2,...,m$)
of $|G\{\theta_j\}\rangle$, she sends
Bob 
$\delta_j\equiv\theta_j+\phi_j'+r_j\pi$,
where $r_j\in\{0,1\}$ is a random binary chosen by Alice
and $\phi_j'$ is the modified version of $\phi_j$ according to
the previous measurement results, which is the standard feed-forwarding
of the one-way model~\cite{Raussendorf}. 
Bob measures $j$th qubit in the basis 
$\{|0\rangle\pm e^{i\delta_j}|1\rangle\}$ and tells the
measurement result to Alice.
\end{itemize}

We call this protocol the single-server protocol,
since there is only a single server (Bob).
It was shown~\cite{BFK} that whatever Bob does he cannot learn
anything about Alice's input, output, and algorithm.

\begin{figure}[htbp]
\begin{center}
\includegraphics[width=0.4\textwidth]{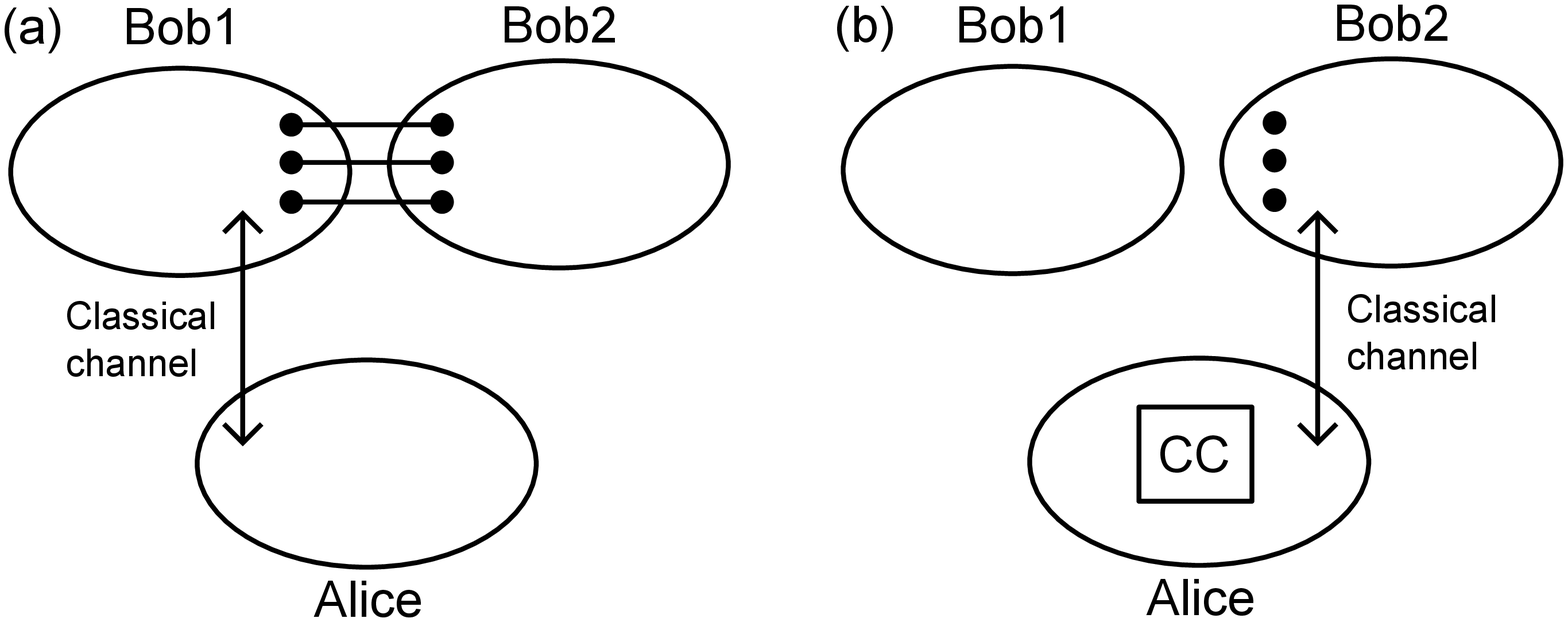}
\end{center}
\caption{
The double-server blind protocol.
(a) Bob1 and Bob2 share Bell pairs.
Alice sends classical messages to Bob1.
Bob1 performs measurements on his qubits of the Bell pairs,
and tells the measurement results to Alice.
(b) Alice and Bob2 run the single-server blind protocol
through the two-way classical channel.
CC is a classical computer.
} 
\label{double}
\end{figure}

In the above single-server protocol, Alice has to have the ability of
emitting randomly-rotated single-qubit states, $\{|\theta_j\rangle\}_{j=1}^m$.
It was shown in Ref.~\cite{BFK} 
that if we have two servers (Bob1 and Bob2) who share
Bell pairs but cannot communicate with each other,
Alice can be completely classical.
(Alice has only to have a classical computer and two classical channels,
one is between Alice and Bob1 and the other is between Alice and Bob2.)
We call such a scheme the double-server scheme,
since there are two servers.
A protocol of the double-server scheme runs as follows~\cite{BFK} (see
also Fig.~\ref{double}):
\begin{itemize}
\item[D1.]
A trusted center distributes Bell pairs to Bob1 and Bob2~\cite{center}.
Now Bob1 and Bob2 share $m$
Bell pairs, $(|00\rangle+|11\rangle)^{\otimes m}$.
\item[D2.]
Alice sends Bob1 classical messages $\{\theta_j\}_{j=1}^m$,
where $\theta_j$ is randomly chosen by Alice
from $\{\frac{k\pi}{4}|k=0,1,...,7\}$.
\item[D3.]
Bob1 measures his qubit of the $j$th Bell pair 
in the basis $\{|0\rangle\pm e^{-i\theta_j}|1\rangle\}$
$(j=1,...,m)$. 
Bob1 tells Alice the measurement results $\{b_j\}_{j=1}^m\in\{0,1\}^m$.
\item[D4.]
After these Bob1's measurements,
what Bob2 has is
$\bigotimes_{j=1}^mZ_j^{b_j}|\theta_j\rangle
=\bigotimes_{j=1}^m|\theta_j+b_j\pi\rangle$.
Now Alice and Bob2 can start the single-server BFK protocol
with the modification 
$\{\theta_j\}_{j=1}^m\to\{\theta_j+b_j\pi\}_{j=1}^m$.
\end{itemize}

In addition to the advantage of the completely classical Alice,
the double-server scheme is intensively studied in
computer science in the context of
the multi-prover interactive proof system,
which assumes computationally unbounded and untrusted prover (server),
and device-independent quantum key distribution~\cite{BFK,Aharonov,Reichardt}.

Note that the impossibility of the communication between two Bobs
is crucial in the double-server protocol. 
If Bob1 can send some message to Bob2,
Bob1 can tell Bob2 $\{\theta_j+b_j\pi\}_{j=1}^m$,
and then Bob2 can learn something about $\{\phi_j\}_{j=1}^m$,
since Bob2 knows $\{\theta_j+b_j\pi+\phi_j'+r_j\pi\}_{j=1}^m$. 
On the other hand, if Bob2 can tell Bob1 $\{\theta_j+b_j\pi+\phi_j'+r_j\pi\}_{j=1}^m$,
Bob1 can learn something about $\{\phi_j\}_{j=1}^m$,
since Bob1 knows $\{\theta_j+b_j\pi\}_{j=1}^m$.
In these cases, the security of Alice's privacy is no longer guaranteed.

In order to perform the correct double-server protocol,
two Bobs must share clean Bell pairs.
Sharing clean Bell pairs is also crucial in many other quantum information
protocols such as the quantum teleportation~\cite{teleportation},
the quantum key distribution~\cite{QKD1,QKD2} and
the distributed quantum computation~\cite{dist1,dist2,dist3,dist4,dist5}.
One standard way of sharing clean Bell pairs in a noisy environment
is the entanglement distillation~\cite{Bennett_PRL,Bennett_PRA,Deutsch,Dur}.
In entanglement distillation protocols,
two people, say Bob1 and Bob2, who want to share clean Bell pairs
start with some dirty $n$ Bell pairs.
Then they perform local operations with
some classical communications,
and finally ``distill" $m$ $(m<n)$ clean Bell pairs~\cite{Bennett_PRL,Bennett_PRA,Deutsch,Dur}.

If we consider the application of the entanglement distillation 
to the double-server blind protocol, 
one huge obstacle is that two Bobs are not allowed to
communicate with each other in the double-server scheme.
Hence, message exchanges between two Bobs,
which are necessary for the entanglement 
distillation, must
be done through the Alice's mediation,
i.e., Bob1 (Bob2) sends a message to Alice, and Alice transfers it
to Bob2 (Bob1).
It is not self-evident that the security of the double-server blind protocol
is guaranteed even if we plug an entanglement distillation protocol
into the double-server blind protocol~\cite{donot}.
For example, Bob1 might send a message to Alice pretending that
it is a ``legal" message for the entanglement distillation.
Alice might naively forward that message to Bob2
without noticing Bob1's evil intention
and believing that it is a harmless message.
In this case, Bob1 can indirectly send some message to Bob2,
and hence the security of the double-server protocol is no longer
guaranteed. 

If the entire entanglement distillation is completed before starting
the double-server protocol, and if Alice delegates her computation to
Bobs only once, then the communication between two Bobs during
the entanglement distillation is harmless, since
when they are doing the entanglement distillation, 
messages related to Alice's computation are not yet sent to Bobs.
However, if Alice delegates more than twice,
then two Bobs might exchange information about the previous
double-server computation during the entanglement distillation
for the next round of the computation as in the case of
the ``device-independence" argument of the quantum key distribution 
with devices having memory~\cite{memory}.
Furthermore, the entanglement distillation might be done in parallel with
the double-server protocol in order to avoid a decoherence.
In these cases, we must be careful about the communication between two Bobs
during the entanglement distillation.
In terms of the composable security, this means that we are interested
in the composable security of the
``distillation $+$ blind computing" protocol~\cite{donot}. 

Throughout this paper, we denote four Bell states
by 
$|\psi_{z,x}\rangle\equiv (I\otimes X^xZ^z)
(|0\rangle\otimes|0\rangle+|1\rangle\otimes|1\rangle)$,
where $(z,x)\in\{0,1\}^2$
and
$X\equiv|0\rangle\langle1|+|1\rangle\langle0|$.

{\it Protocol}.---
Now let us show that the entanglement distillation
by two Bobs is indeed possible without degrading the security.
As in the case of the original BFK double-server protocol,
a trusted center (or Alice) generates $n$ Bell states, $|\psi_{00}\rangle^{\otimes n}$, 
and distribute them to two Bobs;
one qubit of each $|\psi_{00}\rangle$ is sent to Bob1 and
the other to Bob2.
Due to the noise in the channel between the center and Bobs,
each Bell state decoheres, $|\psi_{00}\rangle\to\rho$.
Hence two Bobs share $n$ inpure pairs $\rho^{\otimes n}$,
where $\rho$ is a dirty Bell state: one qubit of $\rho$
is possessed by Bob1 and the other is by Bob2.
Without loss of generality, we can assume that $\rho$
is the Werner state,
$\rho=
F\psi_{11}
+\frac{1-F}{3}(
\psi_{00}
+\psi_{01}
+\psi_{10}
)$,
where $\psi\equiv|\psi\rangle\langle\psi|$.
If it is not the Werner state, it can be converted into the Werner state
by using the twirling operation 
(after applying $I\otimes XZ$)~\cite{Bennett_PRA}.
In order to perform the twirling operation, Alice has only to
randomly choose a $SU(2)$ operator,
and tell its classical description to two Bobs.
Therefore the twirling operation does not affect the security.

Since $\rho$ is Bell-diagonal,
$\rho^{\otimes n}$ is the mixture of tensor products of Bell states:
\begin{eqnarray*}
\rho^{\otimes n}
=
\sum_{(z_1,x_1,...,z_n,x_n)\in\{0,1\}^{2n}}
p(z_1,x_1,...,z_n,x_n)\bigotimes_{j=1}^n\psi_{z_j,x_j}.
\end{eqnarray*}
Alice randomly chooses a $2n$-bit string $s_1$
and sends it to two Bobs.
This $s_1$ is chosen completely randomly being independent of
other parameters (such as $\theta_j$, $\phi_j$, etc.).
Each Bob then performs certain local unitary operation which is
determined by $s_1$.
Each Bob next measures a qubit of a single pair in the computational basis,
and tells the measurement result to Alice.
(The detail of the unitary operation, which is irrelevant here,
is given in Ref.~\cite{Bennett_PRA}.
Which pair is measured is also determined by $s_1$~\cite{Bennett_PRA}.
In brief, these unitary operations and measurements are performed
for obtaining $s_1\cdot v$ (mod2) for the hashing,
where $v\equiv(z_1,x_1,...,z_n,x_n)$.)
From these measurement results by Bobs, Alice can gain a single bit
$s_1 \cdot v$ (mod2) of information.

Since a single pair is measured out, now two Bobs share $n-1$ pairs.
If Alice and two Bobs repeat a similar procedure 
(i.e., Alice randomly chooses a $2(n-1)$-bit string $s_2$ and tells
it to two Bobs. Two Bobs perform local operations, measure a single
pair in the computational basis, and tell the measurement results
to Alice),
Alice can gain another single bit of information.
In this way, they repeat this procedure many times, and 
Alice obtains enough bits
to perform the hashing, which works as follows.

The probability distribution $p(z_1,x_1,...,z_n,x_n)$
has almost all its weight for a set of
$\sim 2^{nS(\rho)}$ ``likely" strings,
where $S(\rho)$ is the von Neumann entropy of $\rho$.
The probability that a $2n$-bit string $(z_1,x_1,...,z_n,x_n)$ 
falls outside of the
set of the $2^{n(S(\rho)+\epsilon)}$ most probable strings
is $O(e^{-\epsilon^2n})$~\cite{Bennett_PRA}.
Therefore, Alice can (almost) specify $p(z_1,x_1,...,z_n,x_n)$
if she gains $nS(\rho)$ bits of information
about $p(z_1,x_1,...,z_n,x_n)$.
This means that it is sufficient for
Alice and two Bobs to repeat the above procedure
for $nS(\rho)$ times. 
Then, two Bobs spend $nS(\rho)$ pairs for measurements, and
therefore at the end of the distillation they share $m\equiv n-nS(\rho)$ pairs,
$\bigotimes_{j=1}^m|\psi_{z_j,x_j}\rangle$, 
where $(z_j,x_j)\in\{0,1\}^2$. Alice
knows the $2m$-bit string $(z_1,x_1,...,z_m,x_m)$. 

After the distillation, Alice and two Bobs can start
the double-server protocol.
Now we modify the double-server protocol as follows:
\begin{itemize}
\item[D1'.]
Two Bobs share
$\bigotimes_{j=1}^m|\psi_{z_j,x_j}\rangle$.
\item[D2'.]
Alice sends Bob1 classical messages
$\{\theta_j'\equiv(-1)^{x_j}\theta_j+z_j\pi\}_{j=1}^m$,
where $\theta_j$ is randomly chosen by Alice from
$\{\frac{k\pi}{4}|k=0,1,...,7\}$.
\item[D3'.]
Bob1 measures his qubit of the $j$th Bell pair
in the basis $\{|0\rangle\pm e^{-i\theta_j'}|1\rangle\}$ ($j=1,...,m$).
Bob1 tells Alice the measurement results $\{b_j\}_{j=1}^m\in\{0,1\}^m$.
\item[D4'.]
The same as D4.
\end{itemize}
Since D4' is the same as D4, it is obvious that
Alice can run the correct single-server blind quantum computation
with Bob2.

{\it Bob1 cannot send any message to Bob2}.---
Let us show that Bob1 cannot send any message to Bob2.
What Bob2 receives from Alice are bit strings, $s_1,...,s_{n-m}$, 
and $\{\theta_j+b_j\pi+\phi_j'+r_j\pi\}_{j=1}^m$.
Since $s_1,...,s_{n-m}$ are completely uncorrelated with
what Bob1 sends to Alice, Bob2 cannot gain any information about
Bob1's message from $s_1,...,s_{n-m}$.
Furthermore,
$r_j$ is randomly taken
by Alice from $\{0,1\}$
being independent of what Bob1 sends to Alice. 
Therefore, Bob2 cannot gain any information about $b_j$
from $\theta_j+b_j\pi+\phi_j'+r_j\pi$.
Bob1 and Bob2 share entangled pairs. However,
due to the no-signaling principle, 
only sharing entangled pairs is useless for message transmission.
Hence Bob1 cannot send any message to Bob2.

{\it Bob2 cannot send any message to Bob1}.---
Next let us show that Bob2 cannot send any message to Bob1.
What Bob1 receives from Alice are
bit strings, $s_1,...,s_{n-m}$,
and $\{\theta_j'\equiv(-1)^{x_j}\theta_j+z_j\pi\}_{j=1}^m$.
Again, $s_1,...,s_{n-m}$, are useless for the message transmission
from Bob2 to Bob1.
Furthermore,
$\theta_j$ is randomly chosen
by Alice from $\{\frac{k\pi}{4}|k=0,1,...,7\}$
being independent of what Bob2 sends to Alice 
and $(z_1,x_1,...,z_m,x_m)$.
Therefore, Bob1 cannot gain any information about Bob2's message
from $\theta_j'$.
Hence Bob2 cannot send any message to Bob1.

{\it Two Bobs cannot learn Alice's computational information}.---
Finally,
let us show the security of Alice's computational information.
First, from Bob2's view point,
the difference between our protocol (i.e., the distillation plus
the modified double-server protocol) and the original BFK double-server
protocol is only that Bob2 receives
bit strings,
$s_1,...,s_{n-m}$, from Alice. 
Since these bit strings are completely uncorrelated
with Alice's computational information,
our protocol is as secure as the original BFK double-server protocol
against Bob2.

Second, from Bob1's view point,
the differences between our protocol and the original BFK
double-server protocol are 
\begin{itemize}
\item[(i)]
Bob1 receives
bit strings, $s_1,...,s_{n-m}$, from Alice.
\item[(ii)]
Bob1 receives 
$\theta_j'\equiv(-1)^{x_j}\theta_j+z_j\pi$
instead of $\theta_j$ from Alice ($j=1,2,...,m$).
\end{itemize}
Again, we can safely ignore (i).
Regarding (ii):
since $\theta_j$ is randomly taken
from
$\{\frac{k\pi}{4}|k=0,1,...,7\}$
being independent of 
Alice's computational information
and $(z_1,x_1,...,z_m,x_m)$,
Bob1 cannot gain any information about Alice's computation from $\theta_j'$.
Hence our protocol is as secure as the original
double-server BFK protocol against Bob1.

TM was supported by JSPS and Program to Disseminate
Tenure Tracking System by MEXT.
KF was supported by MEXT Grant-in-Aid for Scientific Research on
Innovative Areas 20104003.


\begin{thebibliography}{99}
\bibitem{graph}
One might think that Bob can gain some information about
Alice's computation from the graph $G$.
However, there are several ways of hiding Alice's true graph
from Bob~\cite{BFK,FK,topoblind}.
One idea is to use so-called the brick-work state,
which is a graph state with a specific graph where
universal quantum computation can be done with only $X-Y$ plain
measurements~\cite{BFK}.
\bibitem{center}
Since two Bobs cannot communicate with each other,
it is impossible that one Bob generates a Bell pair
and sends a half of it to the other Bob.
\bibitem{donot}
Do not misunderstand that this suggests existing blind protocols
without distillation are insecure:
the stand-alone and composable securities of blind protocols
themselves are already established~\cite{BFK,FK,Vedran,AKLTblind,topoblind,
topoveri,MABQC,composable,composableMA}.
\end{thebibliography}
\end{document}